\documentclass[sigconf, nonacm]{acmart}

\newcommand\vldbdoi{XX.XX/XXX.XX}
\newcommand\vldbpages{XXX-XXX}
\newcommand\vldbvolume{14}
\newcommand\vldbissue{1}
\newcommand\vldbyear{2020}
\newcommand\vldbauthors{\authors}
\newcommand\vldbtitle{\shorttitle} 

\newcommand\vldbavailabilityurl{URL_TO_YOUR_ARTIFACTS}
\newcommand\vldbpagestyle{plain} 

\newcommand{\Hardness}{$\alpha$-Hardness}
\newcommand{\HCBGen}{{HCBGen}}

\newcommand{\acorn}{ACORN}
\newcommand{\nhq}{NHQ}
\newcommand{\ung}{UNG}
\newcommand{\rwalks}{RWalks}

\newcommand{\ra}[1]{\textcolor{black}{#1}}
\newcommand{\rb}[1]{\textcolor{black}{#1}}
\newcommand{\rc}[1]{\textcolor{black}{#1}}
\newcommand{\reviewbox}[1]{}
\usepackage{caption}
\usepackage{makecell}
\usepackage{amsmath}
\usepackage{mathtools}
\usepackage{adjustbox}
\usepackage{algorithm}
\usepackage{algorithmic}
\usepackage{multirow}
\usepackage{xcolor}
\usepackage{tablefootnote}
\usepackage{enumitem}
\usepackage{caption}
\usepackage{etoolbox} %
\usepackage{hyperref} %

\usepackage{todonotes}

\begin{document}
\title{Revisiting Filtered ANN Benchmarks: A Hardness-Controlled Benchmark Generator for Realistic Evaluation}

\settopmatter{authorsperrow=4}

\author{Mintaek Lim}
\affiliation{%
  \institution{Seoul National University}
  \country{South Korea}
  \city{Seoul}
}
\email{victorlim@snu.ac.kr}

\author{Dogeun Kim}
\affiliation{%
  \institution{Seoul National University}
  \country{South Korea}
  \city{Seoul}
}
\email{kdg6245@snu.ac.kr}

\author{Minwoo Kim}
\affiliation{%
  \institution{Seoul National University}
  \country{South Korea}
  \city{Seoul}
}
\email{danny5969@snu.ac.kr}

\author{Jaeyoung Do}
\authornote{Corresponding author.}
\affiliation{%
  \institution{Seoul National University}
  \city{Seoul}
  \country{South Korea}
} 
\email{jaeyoung.do@snu.ac.kr}
\begin{abstract}

Filtered approximate nearest neighbor (FANN) search must satisfy both vector similarity and structured predicates, yet evaluations remain brittle because real hybrid workloads are rarely shareable and existing benchmarks rely on ad-hoc synthetic or semi-real constructions. We argue that realism hinges on execution-driven query difficulty: failures in early filtering trigger over-fetching of additional candidates, shaping latency, throughput, and recall. Building on this insight, we propose \Hardness, a query-level hardness metric that models the conditional execution chain via the over-fetch factor and extends naturally to strategy-conditioned settings. Across diverse datasets and hybrid strategies, \Hardness~ exhibits strong monotonic alignment with empirical performance, while common proxies such as selectivity or attribute–vector correlation are frequently unstable or strategy-inconsistent. We further introduce \HCBGen, a hardness-controlled benchmark generator that uses \Hardness~ as an explicit control signal to synthesize workloads under coarse bias modes or to match a target hardness profile. Our experiments show that widely used benchmarks occupy a narrow, relatively easy portion of the hardness spectrum, masking robustness gaps that emerge under harder queries. Finally, we demonstrate that matching hardness distributions enables privacy-preserving proxy workloads that closely reproduce performance trends, bridging research benchmarks and real evaluation.

\end{abstract}

\maketitle

\pagestyle{\vldbpagestyle}
\begingroup\small\noindent\raggedright\textbf{PVLDB Reference Format:}\\
\vldbauthors. \vldbtitle. PVLDB, \vldbvolume(\vldbissue): \vldbpages, \vldbyear.\\
\href{https://doi.org/\vldbdoi}{doi:\vldbdoi}
\endgroup
\begingroup
\renewcommand\thefootnote{}\footnote{\noindent
This work is licensed under the Creative Commons BY-NC-ND 4.0 International License. Visit \url{https://creativecommons.org/licenses/by-nc-nd/4.0/} to view a copy of this license. For any use beyond those covered by this license, obtain permission by emailing \href{mailto:info@vldb.org}{info@vldb.org}. Copyright is held by the owner/author(s). Publication rights licensed to the VLDB Endowment. \\
\raggedright Proceedings of the VLDB Endowment, Vol. \vldbvolume, No. \vldbissue\ %
ISSN 2150-8097. \\
\href{https://doi.org/\vldbdoi}{doi:\vldbdoi} \\
}\addtocounter{footnote}{-1}\endgroup

\ifdefempty{\vldbavailabilityurl}{}{
\vspace{.2cm}
\begingroup\small\noindent\raggedright\textbf{PVLDB Artifact Availability:}\\
The source code, data, and/or other artifacts have been made available at \url{https://github.com/yoon123seul/hardness_aware_fann_benchmarking.git}.
\endgroup
}

\section{Introduction}

Vector search~\cite{malkov2018hnsw, faiss2019, karpukhin2020dense} has become a first-class primitive in modern data management, powering retrieval-augmented generation (RAG)\cite{lewis2020retrieval}, recommendation engines\cite{covington2016deep, naumov2019deep}, e-commerce search\cite{li2021embeddingbasedproductretrievaltaobao, 10.1145/3397271.3401446}, and multi-modal analytics\cite{radford2021learningtransferablevisualmodels,lu2019vilbertpretrainingtaskagnosticvisiolinguistic}. In production, however, similarity alone is insufficient: queries must also enforce structured predicates such as category, time, or product attributes\cite{dai2024uqequeryengineunstructured, 10.14778/3415478.3415541, rwalks, nhq,485845} that are non-negotiable for correctness and safety. 
\reviewbox{R1.R1, R1.W2}\ra{For example, in ANN-based scholarly search systems, users issue queries in natural language, which are then embedded to retrieve relevant papers based on vector similarity. 
However, users may also require results restricted to specific venues or publication years. 
In such cases, the query must include structured predicates in addition to the embeddings, such as \texttt{"year = 2026 and venue = VLDB"}.}
Without these constraints, the system may return semantically relevant but out-of-scope papers. 
\reviewbox{R1\\W1}\ra{This requires the system to index data containing both structured attributes and unstructured embeddings, and to optimize retrieval so that the returned points satisfy the filter predicates while remaining closest to the query embedding.}

\begin{figure}[!t] 
    \captionsetup{aboveskip=1pt}
    \centering
    \includegraphics[width=0.9\linewidth]{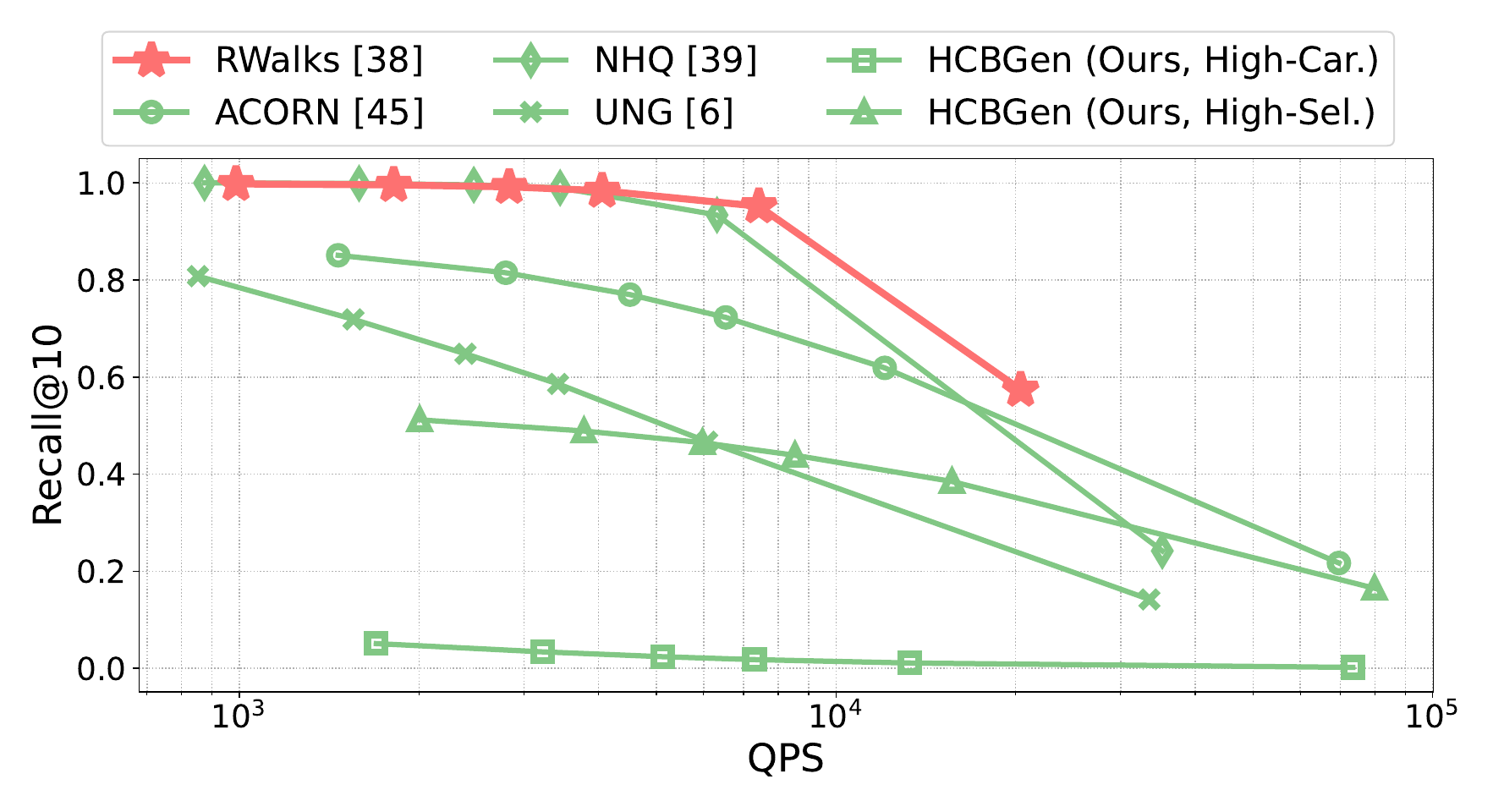}
    \caption{Recall--QPS trade-off curves of the RWalks index evaluated on six synthetic hybrid workloads built from the same vector-only dataset under different label-synthesis policies (four adopted from prior studies and two generated by our hardness-controlled framework: High Cardinality and Highly Selective). Although the workload configuration used in the RWalks paper (red) achieves near-perfect Recall@10 at high throughput, the other workload designs induce substantially different trade-offs, demonstrating that hybrid-index performance can be highly sensitive to workload construction, even with the underlying vector corpus and index fixed.}

    \label{fig:one}
\end{figure}

While algorithmic progress has been rapid, the field faces a quieter but consequential bottleneck: we still lack a reliable and practical way to benchmark hybrid indexing methods~\cite{iff,shi}. Reported results often fail to transfer across datasets, query patterns, and deployment contexts, making it difficult for practitioners to trust which method to adopt and for researchers to draw durable conclusions~\cite{nhq, iff}. The root cause is not that hybrid queries are uncommon, but that real-world hybrid workloads are rarely shareable due to privacy constraints~\cite{privacy2006,privacy2016}, proprietary concerns~\cite{guo2020accelerating, covington2016youtube, dean2013tail}, and limited public coverage~\cite{rwalks}. As a result, most studies evaluate hybrid indices using synthetic or semi-real workloads, typically constructed by taking a vector-only dataset and synthesizing labels, attributes, and filtered queries~\cite{nhq, milvus2021, xu2020mansw}.

Benchmarks therefore exist and are widely used, but they suffer from a fundamental reliability issue: a ``hybrid workload'' is not uniquely determined by the underlying vector dataset. Instead, it depends on benchmark design choices such as how attributes are assigned to vectors, how filters are composed, and how queries are generated, which are often implicit or inconsistent across studies~\cite{shi,acorn, ung}. Even when the base vectors are identical, seemingly minor design changes can substantially alter effective workload difficulty and thus the observed ranking of methods. 
Figure~\ref{fig:one} makes this sensitivity concrete: RWalks~\cite{rwalks}, evaluated on multiple synthetic hybrid workloads built from the same vector-only dataset, SIFT1M~\cite{sift}, under different label-synthesis policies, exhibits widely different recall--QPS trade-offs, implying that without a principled way to characterize and control workload difficulty, hybrid benchmarking may inadvertently favor particular strategies.
Because current synthetic benchmarks neither explicitly model nor sufficiently control query difficulty, such variability is often masked by aggregate reporting based on average-case metrics~\cite{nhq, ung}. In practice, however, real hybrid workloads span a wide range of difficulty, and a small fraction of hard queries can dominate tail latency and expose system-level failures~\cite{wang2024steiner,mybench,pathfinder, dean2013tail, tpch}. Consequently, current benchmarking practices hinder reliable assessment and pose a practical barrier to the adoption of FANN techniques in commercial vector database systems~\cite{pathfinder, boncz2008monetdb, ann-benchmark}.

This challenge stems from the absence of a principled notion of query-level hardness for hybrid search. Existing proxies such as selectivity and correlation~\cite{lin,acorn} capture only partial aspects of difficulty and cannot explain the large performance variance among queries with similar proxy values (Section~\ref{sec:measurements_fail}). As a result, prior work mainly controls dataset properties such as cardinality, skew, correlation, and missingness~\cite{nhq, acorn, ung}, while query workloads are generated through simple heuristics such as random attribute combinations or selectivity targets~\cite{shi}.

To overcome these limitations, we propose \Hardness, a principled definition of query-level hardness for hybrid search, and introduce a hardness-controlled benchmarking methodology in which the construction and evaluation of workloads are directly guided by hardness.
Our key insight is that query difficulty arises from the execution process itself, rather than from isolated static properties. We model FANN execution as a conditional chain, capturing how vector retrieval, over-fetching, and conditional filtering interact and how computational cost accumulates along the execution path~\cite{milvus2021}. This execution-driven formulation yields a unified hardness metric that (i) closely aligns with empirical search performance and (ii) remains robust to previously unmodeled factors, since new influences naturally manifest as changes within the same execution structure rather than requiring ad-hoc proxy extensions.

Hardness is not only diagnostic; it enables a new benchmarking methodology by making workload difficulty explicit and controllable. Building on this model, we introduce a hardness-controlled benchmark generator(\HCBGen) that treats hardness as an explicit control knob for workload construction. Instead of generating queries via ad-hoc rules (e.g., targeting selectivity alone or sampling random filter combinations) the generator systematically synthesizes workloads with prescribed hardness profiles. In particular, it can (i) construct workloads spanning \textsc{Random}-, \textsc{Low}-, and \textsc{High}-hardness regimes for controlled evaluation, and (ii) approximate realistic workloads by matching a target hardness distribution. This design unlocks three practical capabilities: (1) \textbf{Fair comparison:} it evaluates methods under comparable difficulty regimes, reducing the risk that workload construction implicitly advantages a specific strategy. (2) \textbf{Systematic stress testing:} it deliberately instantiates high-hardness workloads to expose failure modes and tail-latency behavior, rather than only average-case performance. (3) \textbf{Privacy-preserving workload sharing:} it enables organizations to share hardness specifications (e.g., hardness distributions) in lieu of raw query logs or sensitive attributes, while still supporting reproducible and predictive performance characterization. Finally, our extensive experiments show that widely used synthetic hybrid benchmarks occupy only a narrow—and relatively easy—portion of the hardness spectrum, highlighting the need for hardness-controlled benchmarking to obtain results that generalize beyond favorable workloads.
Our contributions are as follows:

\begin{itemize}[leftmargin=1.2em]
    \item We define \Hardness, an execution-driven metric for query-level difficulty in FANN that aligns with empirical performance.
    \item We revisit common hybrid benchmarking practices and show that many evaluations are biased toward easy queries, masking tail behavior and limiting generalizability.
    \item We present a hardness-controlled benchmark generator (\HCBGen) that enables controlled workload construction, \ra{fair and index-agnostic benchmarking,} stress testing under targeted hardness regimes, and privacy-preserving approximation of real workloads through hardness specifications.
\end{itemize}

\section{Preliminary}

\subsection{Filtered ANNs}
FANN search extends semantic search by jointly enforcing vector similarity and structured filter constraints.
In standard ANN search, a query retrieves nearest neighbors based only on vector distance.
In contrast, FANN additionally imposes predicates over structured attributes such as categorical labels, numeric ranges, or other metadata constraints~\cite{acorn,nhq}.

Formally, let the base data be defined as
\[
\mathcal{D} = \{ ( \mathbf{v}_i, \mathbf{a}_i ) \mid i = 1, \dots, N \},
\]
where $\mathbf{v}_i \in \mathbb{R}^d$ denotes the $d$ dimensional embedding vector of data point $i$, and $\mathbf{a}_i$ represents its associated set of structured attributes.
A FANN (hybrid) query $q$ is specified by a vector predicate $q_v \in \mathbb{R}^d$ and a set of filter predicates $q_f$, which together define the similarity requirement and the desired attribute constraints of the query.
The goal of FANN search is to retrieve the top-$K$ nearest neighbors to $q_v$ among only those data points satisfing the filter constraint $q_f$:
\[
\operatorname{FANN}(q_{\text{hybrid}})
=
\arg\min_{\substack{S \subseteq \{ i \mid q_f \subseteq \mathbf{a}_i \} \\ |S|=K}}
\ \sum_{i \in S} d(q_v, \mathbf{v}_i).
\]
where $d(\cdot,\cdot)$ denotes a distance function defined on the embedding space.
\reviewbox{R1\\W1}\ra{The additional filter constraints beyond traditional ANN search make FANN fundamentally a two-dimensional retrieval problem, forcing the system to trade off which condition to prioritize during pruning, and thus requiring the index to simultaneously handle unstructured vector similarity and structured evaluation.}

\subsection{Filtered ANN Approaches}
With the growing attention in FANN search, a range of approaches have been proposed, from simple baselines to hybrid-native indexing methods.
The two most common baseline strategies are Post-Filtering and Pre-Filtering.
Post-Filtering retrieves candidates based solely on vector similarity to the query vector $q_v$ and applies the filter predicate afterward.
Because the fraction of filter-satisfying candidates is unknown, this approach typically relies on over-fetching, using a vector-only index such as HNSW~\cite{malkov2018hnsw} followed by filtering.
Pre-filtering applies the filter predicate before vector search, considering only filter-satisfying data points.
While this avoids over-fetching and guarantees full recall, the lack of query-specific indices often forces a brute-force scan over the filtered subset, resulting in poor scalability and query throughput.

Beyond these baselines, several hybrid-native indices have been proposed to more tightly integrate vector similarity and filtering during search, including NHQ\cite{nhq}, ACORN\cite{acorn}, UNG\cite{ung}, RWalks\cite{rwalks}, and Filtered-DiskANN (FDANN)~\cite{filteredDiskANN}.
NHQ extends graph-based ANN search by defining a hybrid distance metric that augments the original vector distance with an attribute-based distance, enabling joint navigation over vector and attribute spaces.
ACORN builds upon HNSW by constructing a denser graph at index build time and restricting traversal at query time to nodes that satisfy the filter predicate, thereby approximating an oracle index tailored to each query.
UNG groups base data points by shared attribute values and constructs a unified navigable graph that explicitly models both intra-group and inter-group connectivity based on attribute inclusion relationships.
RWalks propagates attribute information across the graph during index construction, in a manner reminiscent of node information propagation in graph neural networks~\cite{gnn} to build an attribute matrix, and leverages it at query time to guide search toward filter-satisfying data points.
\reviewbox{R3.R3 R3.W3}\rc{FDANN extends graph-based ANN search with label-aware filtering by starting traversal from entry points guaranteed to contain the query label and exploring only filter-satisfying nodes during search.}

\section{Query-Level Hardness Modeling}
FANN (hybrid) query performance variance is dominated not only by the dataset and index, but also by the \emph{execution dynamics} induced by enforcing both constraints.
We therefore model \emph{query-level hardness} as an execution-grounded quantity that predicts how much \emph{work} a system must perform to return $K$ \emph{valid} neighbors.
This section (i) explains why conventional proxy measures fail to model query hardness, (ii) proposes an execution-grounded hardness model, and (iii) extends the model to be strategy-conditioned.

\subsection{Why Proxy-Based Hardness Fails}
\label{sec:measurements_fail}
Unlike relational database systems\cite{boncz2008monetdb, ioannidis1996query, leis2015how, trummer2017skinnerdb}, query-level hardness in FANN search has been relatively under explored~\cite{lin, zhu2025hybrid}. 
Prior work therefore relies on two intuitive proxy measurements for hybrid query difficulty: {selectivity}, the fraction of base points satisfying the filter, and {correlation}, the degree to which filter-satisfying points align with the query vector's neighborhood~\cite{acorn, lin}. 
In general, lower selectivity tends to increase difficulty because an index must examine more candidates to obtain $K$ valid results, degrading throughput and/or recall~\cite{rwalks}. 
Likewise, higher correlation typically makes queries easier: when filter-satisfying points are spatially clustered near the query vector, graph-based indices are more likely to reach valid targets through graph paths, improving recall while shrinking the explored search space and increasing throughput.

However, selectivity and correlation only partially describe the \emph{execution} of FANN. 
Both are one-dimensional, static statistics, whereas actual difficulty is shaped by execution dynamics (i.e., how candidate retrieval, over-fetching, and filtering interact until $K$ valid neighbors are obtained.) 
Consequently, the same global selectivity can induce very different \emph{local} availability of filter-satisfying points around the query, and correlation do not directly determine how many candidates must be explored before $K$ valid results are found. 
As a result, the ordering implied by selectivity or correlation can differ substantially from the ordering induced by empirical recall--QPS trade-offs.

\begin{figure}[t]
    \centering
    \includegraphics[width=1.0\linewidth]{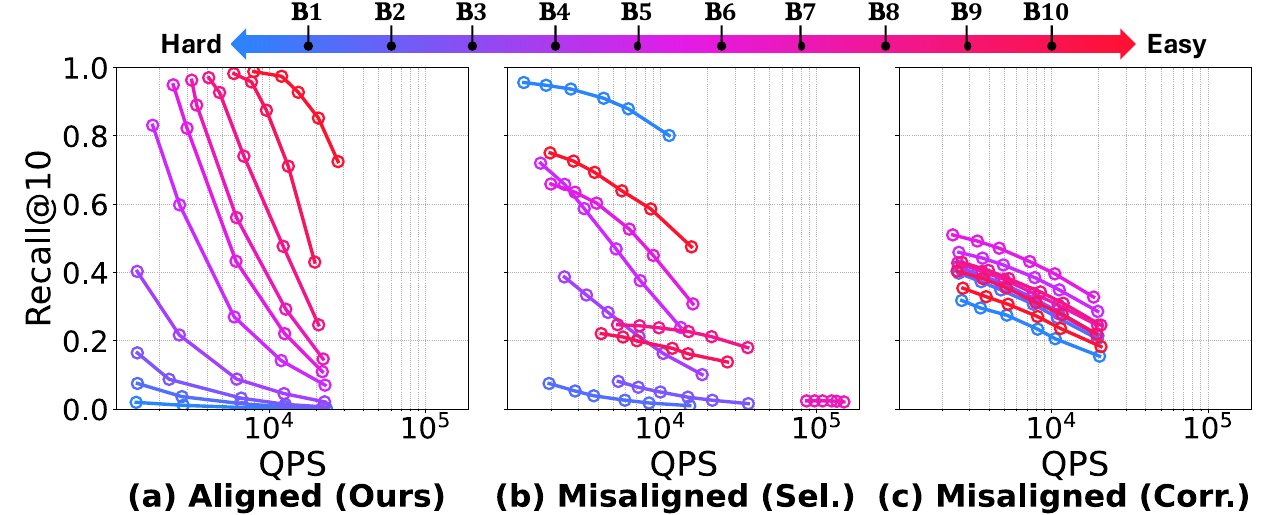}
    \caption{Success and failure cases of query ordering for recall–QPS trade-offs. For each proxy, queries are sorted by the estimated score and split into 10 equal-sized percentile bins (B1=hardest 10\% → B10=easiest 10\%). We plot recall–QPS curves for each bin on a synthetic SIFT1M-based benchmark using RWalks: (a) Successfully aligned (Ours), (b) Misaligned (Selectivity), and (c) Misaligned (Correlation).}

    \label{fig:measurements_fail}
\end{figure}

Figure~\ref{fig:measurements_fail} illustrates representative failure cases of this misalignment. 
Using SIFT1M~\cite{sift}, we synthesize a hybrid workload by assigning three categorical attributes (cardinality 6 each) and generating 10{,}000 queries (details in Section~\ref{sec:hardness_validation}). 
We sort queries by each proxy, partition them into 10 bins, and evaluate each bin independently using RWalks~\cite{rwalks}. 
A lower bin index corresponds to a lower proxy value (interpreted as ``harder''), while curves closer to the upper-right indicate better recall--QPS trade-offs (``easier''). 
(a) is included as a reference case (sorted by \Hardness, Section~\ref{sec:hardness_validation}), illustrating a setting where bin-wise query performance is well aligned and exhibits a consistent ordering.
Yet, in both (b) and (c), neither selectivity nor correlation yields a consistent ordering that aligns with actual performance. 
These observations indicate that query hardness cannot be reliably characterized by static, measurement-based proxies alone.

More broadly, a straightforward attempt to define a single scalar hardness by composing multiple measurements suffers from two fundamental limitations. 
First, such measurements provide only indirect signals and do not explicitly model search execution, so alignment with performance is not guaranteed. 
Second, measurement-based definitions are inherently incomplete: newly discovered hardness-related factors require revising both the hardness definition and the associated measurement set. 
This motivates an execution-grounded hardness definition that is anchored in the search process rather than in ad-hoc proxy statistics.

\subsection{Execution-Grounded Hardness via Conditional Over-Fetching}
\label{sec:conditional_chain}

To overcome the limitations of measurement-based proxies, we define query-level hardness by explicitly modeling FANN execution as a \emph{conditional chain}.
The key challenge in hybrid search is that the system must jointly satisfy vector similarity and a structured predicate; when early-stage filtering fails to produce enough valid results, the search must \emph{over-fetch} and continue exploring, a behavior commonly observed in practical systems (e.g., post-filtering pipelines)~\cite{milvus2021, nhq}. 
Our formulation captures this execution dynamic through an over-fetch factor, yielding (i) stronger alignment with empirical performance and (ii) a structurally complete definition whose functional form does not require revision when new hardness-related phenomena are discovered.
\vspace{0.1cm}

\indent\textit{Query and execution setting.} A hybrid query is $q=(q_v, q_f)$ where $q_v$ is the query vector and $q_f$ is a set of structured predicates.
Let $K$ be the target number of returned neighbors.
Let $V_0$ denote the full base data and $V_f \subseteq V_0$ the subset satisfying $q_f$.

\vspace{0.1cm}
\textit{Key idea.} Hybrid query difficulty is governed by \emph{how many vector-ranked candidates must be examined before execution outputs $K$ filter-satisfying results}.
To make this execution-grounded notion concrete, we start from a canonical execution chain in which over-fetching is explicit (post-filtering), and then extend the model to be strategy-conditioned in Section~\ref{sec:strategy_dependent}.

\subsubsection{A canonical conditional execution chain}

We instantiate query-level hardness by tracing a simple yet representative
\emph{post-filtering} execution chain:
(1) retrieve candidates according to vector similarity;
(2) scan candidates and keep only those satisfying $q_f$ until $K$ valid neighbors are obtained.
This chain is methodological rather than restrictive: it is the simplest setting in which
over-fetching is explicit, making it a clean basis for defining and estimating hardness,
and it matches the vector-centric execution behavior studied throughout this section.


\begin{definition}[Over-fetch factor]
\label{def:alpha}
For a hybrid query $q=(q_v,q_f)$ and target $K$,
the over-fetch factor $\alpha(q;K)$ is the number of vector-ranked candidates that must be examined
so that the scan phase outputs $K$ results satisfying $q_f$.
\end{definition}

\subsubsection{\Hardness~ as conditional execution cost}

We define \Hardness~ \\ as the \emph{execution cost} induced by the above conditional chain. To capture the conditional dependence between stages, namely that over-fetching amplifies not only predicate check work but also the required candidate retrieval effort under a fixed operating point, we use a multiplicative form.

Let $H_{\text{fetch}}(q_v \mid m)$ denote the cost of generating the top-$m$
vector-ranked candidates for $q_v$ from $V_0$. Let $H_{\text{scan}}(q_f \mid q_v, m, K)$ denote the cost of scanning these $m$ candidates and extracting $K$ filter-satisfying results. Then the \Hardness~ of $q$ is defined as
\begin{equation}
H_{\alpha}(q|K) \triangleq H_{\text{fetch}}(q_v|\alpha(q;K)) \times H_{\text{scan}}(q_f|q_v, \alpha(q;K), K).
\label{eq:hpost_mult}
\end{equation}
Under post-filtering, predicate checking incurs an (approximately) constant per-candidate cost,
so scan hardness scales linearly with the number of examined candidates:
\begin{equation}
H_{\text{scan}}(q_f \mid q_v, m, K) = C_{\text{scan}}\cdot m.
\label{eq:hscan_linear}
\end{equation}
Plugging $m=\alpha(q;K)$ into Eq.~\eqref{eq:hscan_linear} and substituting into Eq.~\eqref{eq:hpost_mult} yields
\begin{equation}
H_{\alpha}(q \mid K)
\;\propto\;
H_{\text{fetch}}(q_v \mid \alpha(q;K)) \cdot \alpha(q;K).
\label{eq:hpost_simplified}
\end{equation}

\indent\textit{Monotonicity and alignment with performance.} Hardness should increase monotonically with chain cost: larger $\alpha$ raises scanning cost and drives retrieval deeper, increasing $H_{\text{fetch}}(q_v\mid \alpha)$. Thus, higher $H_{\alpha}$ implies lower throughput and/or recall under a fixed budget. Equivalently, on a given dataset $\mathcal{D}$, $H_{\alpha}$ is designed to be positively associated with $\left(\frac{1}{\mathrm{Recall}}\cdot\frac{1}{\mathrm{QPS}}\right)$, as validated in Section~\ref{sec:hardness_validation}.
\vspace{0.1cm}
\indent\textit{Reducing $H_{\text{fetch}}$ to vector-only difficulty.}
Conditioned on the required depth $m=\alpha(q;K)$, the fetch component depends only on the vector predicate, reducing to vector search up to depth $m$.
\reviewbox{R3.R1 R3.W1}\rc{Thus, $\alpha$-Hardness generalizes Steiner-Hardness~\cite{wang2024steiner} from vector-only to hybrid search: $H_{\text{fetch}}$ captures the Steiner-Hardness of the full graph at depth $\alpha(q;K)$, while the overall formulation also accounts for filter-induced over-fetching.}
Since $\alpha(q;K)$ is not known \emph{a priori}, we estimate it with $\widehat{\alpha}(q;K)$ and substitute $\widehat{\alpha}(q;K)$ into $H_{\alpha}$.

\subsubsection{Estimating $\widehat{\alpha}(q;K)$ via selectivity and local availability}
\label{sec:alpha_estimation}

\reviewbox{R2\\W1}\rb{Semantically, $\widehat{\alpha}(q;K)$ represents how many vectors must be traversed around the query vector $q_v$ in order to obtain $K$ filter-satisfying results.
A natural first signal for this quantity is the \emph{global selectivity} of the filter, $s = |V_f|/|V_0|$.
When this fraction is small, valid results are rare in the dataset as a whole, and $\widehat{\alpha}$ is therefore likely to increase.
However, global selectivity alone is insufficient because it ignores how $V_f$ is distributed around $q_v$.
Vectors and labels can be correlated; for example, papers from a specific year may cluster around certain topics in the embedding space.
As a result, even when a filter is globally selective, the filter-satisfying vectors may still be concentrated near the query, in which case the effective over-fetching required to collect $K$ valid results can remain small.}

To capture this effect, we estimate the over-fetching needed to obtain $K$ valid results using both \emph{global selectivity} and a \emph{local availability (density) correction}:
\begin{equation}
\widehat{\alpha}(q;K)
\;\triangleq\;
\frac{1}{s}\cdot \frac{1}{\rho(q)} \cdot K,
\label{eq:alpha_hat_simple}
\end{equation}
where $\rho(q)$ quantifies how much denser filter-satisfying points are around $q_v$ compared to the full data.

To operationalize local availability, we use a query-local density proxy based on the ratio of $K$-NN distances in $V_0$ and $V_f$:
\begin{equation}
\widehat{\alpha}(q;K)
=
\frac{|V_0|}{|V_f|}
\cdot
\frac{
d\!\left(q_v, K\text{-NN}\mid V_f\right)
}{
d\!\left(q_v, K\text{-NN}\mid V_0\right)
}
\cdot K,
\label{eq:alpha_hat}
\end{equation}
where $d(q_v, K\text{-NN}\mid V)$ denotes the distance from $q_v$ to its $K$-th nearest neighbor within set $V$.
Eq.~\eqref{eq:alpha_hat} increases $\widehat{\alpha}$ when (i) the filter is globally selective (small $|V_f|$) and/or
(ii) valid points are locally sparse near the query (large $d(\cdot\mid V_f)$ relative to $d(\cdot\mid V_0)$).

Finally, substituting $\widehat{\alpha}(q;K)$ into Eq.~\eqref{eq:hpost_simplified} yields a practical hardness score that can be computed
\emph{without running} the target hybrid index, while remaining tightly coupled to the execution chain.
We empirically validate that ordering queries by $H_{\alpha}$ aligns well with recall--QPS trends
(Section~\ref{sec:hardness_validation}).

\vspace{0.1cm}
\textit{Completeness and extensibility.}
This formulation supports a \emph{structurally complete} notion of hybrid query hardness: $H_{\alpha}$ is determined by the execution chain and its cost components, rather than by an open-ended set of ad-hoc proxies.
Thus, new hardness-related phenomena can be incorporated by refining estimators of the existing chain variables, most notably $\widehat{\alpha}(q;K)$, within the same framework.
Our results in Section~\ref{sec:hardness_validation} show that this estimation already captures the dominant sources of over-fetching in practice and aligns well with empirical performance.

\subsubsection{Extending the Execution Chain to Complex Predicates}
\label{sec:complex_predicates}

\reviewbox{R3\\E6}\rc{
Eq.~\eqref{eq:hscan_linear} is sufficient for the categorical-predicate regime targeted by existing FANN indices (e.g., UNG, RWalk, NHQ, and FDANN), where treating per-candidate predicate-check cost as approximately constant is appropriate.
However, real applications may involve more complex predicates supported by systems such as pgvector~\cite{pgvector}, including polygon and regex matching.
To accommodate future indices that may support such queries, we show that Eq.~\eqref{eq:hscan_linear} can be naturally extended to cases where predicate-evaluation cost varies substantially across queries.
The framework handles this by refining the scan-cost term while preserving the hardness formulation:
\begin{equation}
H_{\mathrm{scan}}\!\left(q_f \mid q_v, \widehat{\alpha}(q;K), K\right)
\;=\;
\widehat{\alpha}(q;K)\cdot c_{\mathrm{scan}}(q_f),
\label{eq:hscan_complex}
\end{equation}
where $c_{\mathrm{scan}}(q_f)$ is the average per-candidate predicate-evaluation cost.
A natural refinement is to write
\begin{equation}
c_{\mathrm{scan}}(q_f) = c(q_f)\cdot \phi(q_f),
\label{eq:cscan_decompose}
\end{equation}
where $c(q_f)$ is a predicate-type-specific unit cost, calibrated empirically~\cite{costmodel}, and $\phi(q_f)$ captures predicate complexity in terms of required unit operations~\cite{measuring}.
For example, $\phi(q_f)$ may correspond to the number of vertices in a polygon predicate~\cite{polygon} or the number of wildcards in a regular-expression predicate~\cite{regex1, regex2}.
For simple categorical predicates, $c_{\mathrm{scan}}(q_f)$ is approximately constant, recovering Eq.~\eqref{eq:hscan_linear}.
This natural extension preserves the main advantage of the execution-grounded framework: richer hardness-related phenomena can be incorporated by refining existing chain variables, rather than by introducing an unrelated proxy. Experimental validation of this extension is provided in Section~\ref{sec:complex_predicate_results}.}

\begin{figure*}[t!]
    \centering
    \includegraphics[width=1\linewidth]{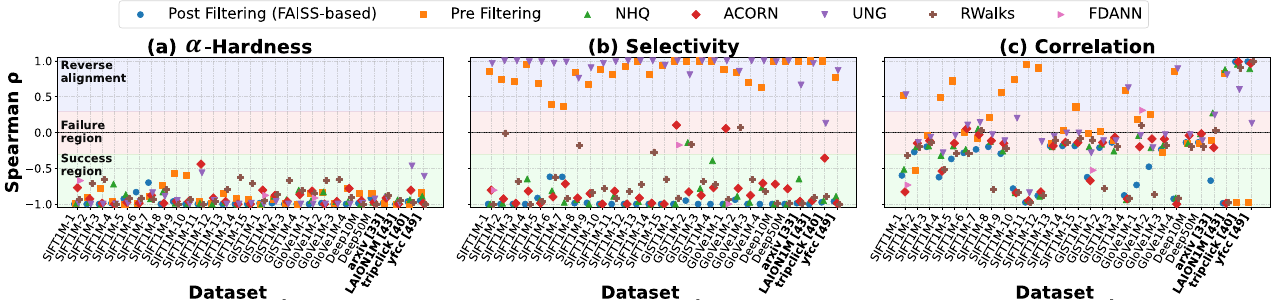}
    \caption{Spearman rank correlation between query hardness estimates (\Hardness~(ours) or baseline proxies: selectivity and correlation) and search performance across 29 datasets (25 synthetic and 4 semi-real), evaluated under seven hybrid query processing strategies. Semi-real workloads are constructed by taking a vector-only dataset and synthesizing labels, attributes, and filtered queries. Synthetic datasets are identified using the configuration IDs (Table~\ref{tab:dataset_configurations}). Values closer to $-1$ indicate stronger alignment (harder queries $\rightarrow$ worse performance), whereas values in $[0.3, 1]$ indicate unintended reverse alignment.}
    \label{fig:total_heatmap}
\end{figure*}

\subsection{Strategy-Conditioned Hardness Estimation}
\label{sec:strategy_dependent}

Although $H_{\alpha}$ is defined by modeling the post-filtering execution chain from a vector-centric viewpoint, existing FANN approaches are not limited to this family.
As discussed in prior surveys on FANN indices~\cite{lin}, existing methods can be broadly categorized into \emph{vector-centric} and \emph{filter-centric} pruning families\footnote{\ra{Although indices such as UNG and ACORN incorporate both filter and vector predicates through hybrid pruning mechanisms, they can still be classified into one of the two families according to the dominant pruning behavior. This categorization is also consistently supported by our experimental results.}}.
Filter-centric strategies do not follow the post-filtering chain; instead, they first restrict the search space using the filter predicate and then perform vector search within the filtered subset.
Because hybrid-query difficulty is defined through the execution chain, it is inherently \emph{strategy-dependent}: different indices enforce vector and filter constraints in different orders, leading to different dominant cost drivers.
Accordingly, we denote the resulting score as $H(q\mid \text{strategy})$.

\vspace{0.1cm}
\reviewbox{R2\\W2}\rb{\textit{Pre-filtering execution chain.}
For filter-centric approaches such as UNG, FANN execution can be modeled as first inducing a filter-satisfying subgraph from the full graph index, and then performing vector search within that subgraph.
Let $G_0$ denote the original graph over the full dataset $V_0$, and let $G_f$ denote the induced subgraph over $V_f$.
The corresponding execution chain is
\[
H_{\mathrm{PreFiltering}}(q\mid K)
\triangleq
H_{\mathrm{restrict}}(q_f\mid G_0)\cdot H_{\mathrm{search}}(q_v\mid G_f),
\]
where $H_{\mathrm{restrict}}$ denotes the cost of identifying the filter-satisfying search space and $H_{\mathrm{search}}$ the cost of vector search within $G_f$.
\reviewbox{R3.R1 R3.W1}\rc{This hardness can be interpreted as an execution-based approximation to the Steiner-Hardness~\cite{wang2024steiner} of the filter-induced subgraph $G_f$, extending the same perspective in Section~\ref{sec:alpha_estimation} to the filter-centric setting.}
Since the restriction step is relatively stable for a fixed index, and search over $G_f$ degrades toward a linear scan as selectivity decreases, the dominant cost factor is the size of the filtered vectors:
\[
H_{\mathrm{PreFiltering}}(q\mid K)\propto H_{\mathrm{search}}(q_v\mid G_f)\propto |V_f|.
\]
By rearranging Eq.~\eqref{eq:alpha_hat_simple}, we obtain
\[
|V_f|
=
\frac{|V_0|\,K\,\rho(q)}{\widehat{\alpha}(q;K)}.
\]
Pre-filtering already restricts search to the filter-satisfying subgraph $G_f$, so the dominant family-level cost is governed mainly by the size of that restricted search space, $|V_f|$, rather than by the local-density term $\rho(q)$. 
Since $\rho(q)$ governs how quickly post-filtering accumulates valid results during vector-centric traversal—it ceases to be a primary cost driver once search is already confined to $V_f$.
With $K$ and $|V_0|$ fixed for a given benchmark setting, the dominant filter-centric cost follows the opposite monotone trend from vector-centric cost and decreases as the effective $\widehat{\alpha}$ increases.
}

\vspace{0.1cm}
\textit{Why a simple inversion is sufficient.}
The above derivation establishes a monotone duality between the two execution families: queries that are hard for vector-centric execution tend to be easy for filter-centric execution, and vice versa.
Rather than introducing separate hardness models per family, we therefore apply a monotone inversion to reuse the same
execution-grounded score:
\[
H(q\mid \text{strategy})
\triangleq
\begin{cases}
H_{\alpha}(q), & \text{if the strategy is vector-centric}, \\
\dfrac{1}{H_{\alpha}(q)}, & \text{if the strategy is filter-centric}.
\end{cases}
\]
This transformation is intentionally conservative: it only assumes that the two strategy families exhibit opposite monotone trends
with respect to over-fetching-driven difficulty (as observed empirically), without committing to a specific internal implementation.
As shown in Section~\ref{sec:hardness_validation}, although this minimal strategy-aware transformation is not intended as an exact cost model for the filter-centric implementations, it is sufficient to achieve strong alignment with empirical performance across both families.

\subsection{Hardness Validation}
\label{sec:hardness_validation}

We empirically validate whether the proposed \Hardness~ induces a query ordering that is consistent with observed \emph{recall--throughput} trade-offs in FANN search. Our goal is not to fit a particular index, but to test whether the hardness definition remains predictive across (i) diverse dataset/workload characteristics and (ii) fundamentally different hybrid execution strategies.

\subsubsection{Dataset \& Workload Suite}
\label{sec:dataset_eval_method}

Following prior hybrid benchmarking practice~\cite{ung,nhq,acorn}, we construct a suite of \emph{synthetic} hybrid datasets by augmenting standard vector-only benchmarks with synthesized categorical labels, while systematically controlling the core label/workload factors: the source base vectors, number of attributes, attribute cardinalities, value distributions, per-attribute missing probabilities, and \emph{data--label correlation (DLC)} between base vectors and base labels\footnote{DLC is distinct from the query--vector correlation (QVC) used as a hardness proxy in Section~\ref{sec:measurements_fail}; DLC characterizes correlation in the \emph{data/label generation} process, whereas QVC is measured at the \emph{query level}}.
 
Query workloads are generated using the same synthesis policy used for the base data to ensure structural compatibility. All synthetic configurations are summarized in Table~\ref{tab:dataset_configurations}. 
In addition, to verify that \Hardness~ is not an artifact of fully synthetic construction, we include four \emph{semi-real} datasets whose base data—base vectors paired with corresponding labels—are drawn from real-world corpora, and whose query workloads are sampled or synthetically generated from the base data 
~\cite{arxiv, laion, tripclick, yfcc}.

We evaluate alignment under seven representative hybrid query processing strategies:
FAISS\cite{faiss2019}-based Post-Filtering and Pre-Filtering, and five hybrid-native indices (NHQ~\cite{nhq}, ACORN~\cite{acorn}, UNG~\cite{ung}, RWalks~\cite{rwalks}), and \reviewbox{R3.R3 R3.W3}\rc{FDANN}~\cite{filteredDiskANN}\footnote{Since FDANN can handle only queries with a single filter attribute, we report results only for SIFT1M-1, GIST1M-1, and GloVe1M-1.}.)
For each method, we sweep its parameters to obtain recall--QPS trade-offs, following the recommended settings in the corresponding original papers
.
\rc{\reviewbox{R3.R2 R3.W2}To further examine whether the observed behavior remains robust when the dataset size exceeds CPU cache capacity, we additionally include the Deep10M and Deep50M datasets, which are subsets of Deep1B~\cite{7780595}.} For additional experimental details refer to Section~\ref{sec:experiments_setting}.

\begin{table}[t]
\footnotesize
\centering
\caption{
Parameter configurations for synthetic datasets. The configuration ID encodes the source base vectors (prefix). For brevity, a single value indicates a setting applied uniformly to all attributes; mixed settings list per-attribute values.}
\label{tab:dataset_configurations}
\setlength{\tabcolsep}{2pt}
\renewcommand{\arraystretch}{0.8}
\begin{adjustbox}{max width=1\linewidth} 

\begin{tabular}{c c c c c c}
\toprule
ID & \#Attr& Cardi.& Distribution & Missing Prob.& DLC \\
\midrule

SIFT1M-1  & 1  & 12 & \multirow{7}{*}{Zipf}& \multirow{7}{*}{0.5}& \multirow{7}{*}{0.0}\\
SIFT1M-2  & 3  & 6  &                      &                      &                      \\
SIFT1M-3  & 3  & 12 &                      &                      &                      \\
SIFT1M-4  & 12 & 1  &                      &                      &                      \\
SIFT1M-5  & 12 & 3  &                      &                      &                      \\
SIFT1M-6  & 12 & 6  &                      &                      &                      \\
SIFT1M-7  & 12 & 12 &                      &                      &                      \\

\midrule

SIFT1M-8  & 3  & 12 & \multirow{2}{*}{Random} & \multirow{2}{*}{0.5} & \multirow{2}{*}{0.0} \\
SIFT1M-9  & 12 & 3  &                         &                      &                      \\

\midrule

SIFT1M-10  & \multirow{3}{*}{3} & \multirow{3}{*}{12} & \multirow{3}{*}{Zipf} & \multirow{3}{*}{0.5} & 0.5 \\
SIFT1M-11 &                    &                     &                       &                      & 1.0 \\
SIFT1M-12 &                    &                     &                       &                      & mix \\

\midrule

SIFT1M-13 & \multirow{3}{*}{3} & \multirow{3}{*}{12} & \multirow{3}{*}{Zipf} & 0.0 & \multirow{3}{*}{0.0} \\
SIFT1M-14 &                    &                     &                       & 0.8 &                      \\
SIFT1M-15 &                    &                     &                       & mix &                      \\

\midrule

GIST1M-1  & 1                  & 12                  & Zipf   & 0.5 & 0.0 \\
GIST1M-2  & 3                  & 12                  & Random & 0.5 & 0.0 \\
GIST1M-3  & 3                  & 12                  & Zipf   & 0.5 & 0.0 \\
GIST1M-4  & 3                  & 12                  & Zipf   & 0.0 & 0.5 \\

\midrule

GloVe1M-1 & 1                  & 12                  & Zipf   & 0.5 & 0.0 \\
GloVe1M-2 & 3                  & 12                  & Random & 0.5 & 0.0 \\
GloVe1M-3 & 3                  & 12                  & Zipf   & 0.5 & 0.0 \\
GloVe1M-4 & 3                  & 12                  & Zipf   & 0.0 & 0.5 \\

\midrule

\rc{Deep10M}   & \rc{3}                  & \rc{12}                  & \rc{Zipf}   & \rc{0.5} & \rc{0.0} \\
\rc{Deep50M}   & \rc{3}                  & \rc{12}                  & \rc{Zipf}   & \rc{0.5} & \rc{0.0} \\

\bottomrule
\end{tabular}
\end{adjustbox}
\end{table}

\subsubsection{Alignment Protocol and Metrics}
FANN search performance is commonly summarized by recall--QPS trade-offs under varying parameters~\cite{faiss2019,shi}. Accordingly, a hardness measure is \emph{well-aligned} if sorting queries by hardness yields groups whose trade-off curves degrade monotonically. We use Recall@10 and refer to it simply as recall throughout this section.
For each dataset and strategy, we compute $H(\mathbf{q}\mid\text{strategy})$ (Section~\ref{sec:strategy_dependent}), sort queries by increasing hardness, and partition them into $B{=}10$ equal-sized batches.
This operationalizes the ordering soundness requirement: later batches should contain queries that are no easier than earlier ones\footnote{We use $B{=}10$ to balance resolution and stability; our conclusions are unchanged for nearby choices such as 5 or 20.}.
For each batch, we evaluate the target strategy across operating points, yielding one recall--QPS trade-off curve. To summarize each curve while preserving the trade-off, we compute a Pareto-optimal score by extracting the Pareto frontier, averaging recall and QPS over frontier points, and taking their product~\cite{pareto}. Higher scores indicate more efficient query processing.

Let $\overline{H}_b$ be the average hardness of batch $b$, and let $\textsc{Score}_b$ be the Pareto-optimal score of that batch. We quantify alignment by the Spearman rank correlation between $\{\overline{H}_b\}_{b=1}^{B}$ and $\{\textsc{Score}_b\}_{b=1}^{B}$~\cite{spearman1904proof}. Because higher hardness should correspond to lower performance, stronger alignment is reflected by Spearman's $\rho$ closer to $-1$. For conventional hardness proxies (selectivity and correlation), we negate them before batching so that larger values consistently indicate higher difficulty, matching the direction of $H(\cdot)$\footnote{We intentionally do \emph{not} choose proxy directions per strategy, since that would leak strategy knowledge into the baseline and artificially improve its alignment.}.

\begin{table}[t]
\centering
\footnotesize
\setlength{\tabcolsep}{5pt}
\renewcommand{\arraystretch}{1.1}
\caption{\rc{Alignment comparison between Extended and Original \Hardness~ across predicate workloads. Predicate complexity is grouped by calibrated unit scan cost $c(q_f)$. Improvement is measured as $\Delta |\rho| = |\rho_{\textsc{Adaptive}}| - |\rho_{\textsc{Original}}|$.}}
\label{tab:complex_predicate_raw}
\begin{adjustbox}{max width=\columnwidth}
\begin{tabular}{l l c c c c}
\toprule
Complexity & Dataset & $c(q_f)$ & $\rho$ (Original) & $\rho$ (Extended) & $\Delta |\rho|$ \\
\midrule
\multirow{2}{*}{Simple}
& categorical  & 1.56   & -0.99 & -0.99 & +0.00 \\
& range        & 2.18   & -0.96 & -0.92 & -0.04 \\
\midrule
\multirow{4}{*}{Medium}
& p-simple     & 3.75   & -0.82 & -0.98 & +0.16 \\
& p-medium     & 3.75   & -0.62 & -1.00 & +0.38 \\
& p-complex    & 3.75   & -0.59 & -1.00 & +0.41 \\
& set-overlap  & 10.76  & -0.90 & -1.00 & +0.10 \\
\midrule
\multirow{2}{*}{Complex}
& r-simple     & 287.92 & -0.89 & -0.99 & +0.10 \\
& r-complex    & 287.92 & -0.66 & -0.95 & +0.29 \\
\bottomrule
\end{tabular}
\end{adjustbox}
\end{table}

\subsubsection{Results and Takeaways}
Figure~\ref{fig:total_heatmap} reports Spearman's $\rho$ across datasets and strategies.
Across diverse dataset/strategy combinations, the proposed hardness exhibits strong monotonic alignment with empirical search performance (typically $\rho \le -0.7$), whereas proxy measurements are frequently near zero or unstable, indicating that they do not reliably order query difficulty.

The baseline \emph{query--vector correlation (QVC)} becomes informative only when strong positive \emph{data--label correlation (DLC)} is explicitly enforced in the synthetic construction (e.g., SIFT1M-10, -11, -12, GIST1M-4, and GloVe1M-4),
a condition that is neither universal nor guaranteed in realistic workloads; outside such regimes, QVC provides limited explanatory power.

Selectivity exhibits opposite signs of $\rho$ across strategy families because it is intrinsically entangled with the pruning order, leading to reversed alignment with the intended ordering under certain strategies. 
For vector-centric strategies, lower selectivity typically makes queries harder by increasing over-fetching and predicate checks, whereas for filter-centric strategies it can make queries easier by shrinking the pre-filtered search space and reducing downstream vector traversal.
Without explicit knowledge of the execution family, which is typically unavailable to a benchmark designer, selectivity cannot even determine a \emph{consistent direction} of difficulty, and in many settings yields weak alignment.

If we treat $\rho \in [-0.3, 0.3]$ as an alignment failure case~\cite{cohen1988statistical}, \Hardness~ incurs no such failures in any of our evaluated settings, \rc{including scaled-out regimes,} whereas selectivity fails in a non-trivial fraction of cases. This gap arises because selectivity isolates only the filter predicate while ignoring the vector predicate and their joint interaction along the execution chain, thereby violating the validity and ordering soundness criteria discussed in Section~\ref{sec:conditional_chain}.

\subsubsection{Validation on Complex Predicates}
\label{sec:complex_predicate_results}

\reviewbox{R3\\E6}\rc{
We further validate the extended formulation(Eq.~\eqref{eq:hscan_complex}) on complex-predicate workloads built on SIFT100K, a randomly sampled subset of SIFT1M. The evaluated predicates include range filters, polygon matching with 4--12 (p-simple), 30--70 (p-medium), or 150--250 (p-complex) vertices, regex matching over strings of length 100 with either no wildcard (r-simple) or 1--8 wildcards (r-complex), and set-overlap predicates with a vocabulary pool of size 100. Since no existing hybrid index directly supports this predicate set, we use HNSW-based post-filtering as the representative execution strategy.}

\rc{
Table~\ref{tab:complex_predicate_raw} shows that the original \Hardness~ aligns well when predicate complexity is low, but its ordering quality degrades as per-candidate validation cost becomes more query-dependent, especially for polygon and regex workloads.
By contrast, the extended formulation remains stably aligned across all workloads, with $\rho < -0.9$ throughout, and yields the largest gains on structurally complex predicates such as \texttt{p-medium}, \texttt{p-complex}, and \texttt{r-complex}.
These results show that \Hardness~ extends naturally to compute-bound settings, where the dominant cost shifts from over-fetching to candidate scanning, without changing the underlying conditional chain.
}

\begin{figure}[t!]
    \centering
    \includegraphics[width=1.02\linewidth]{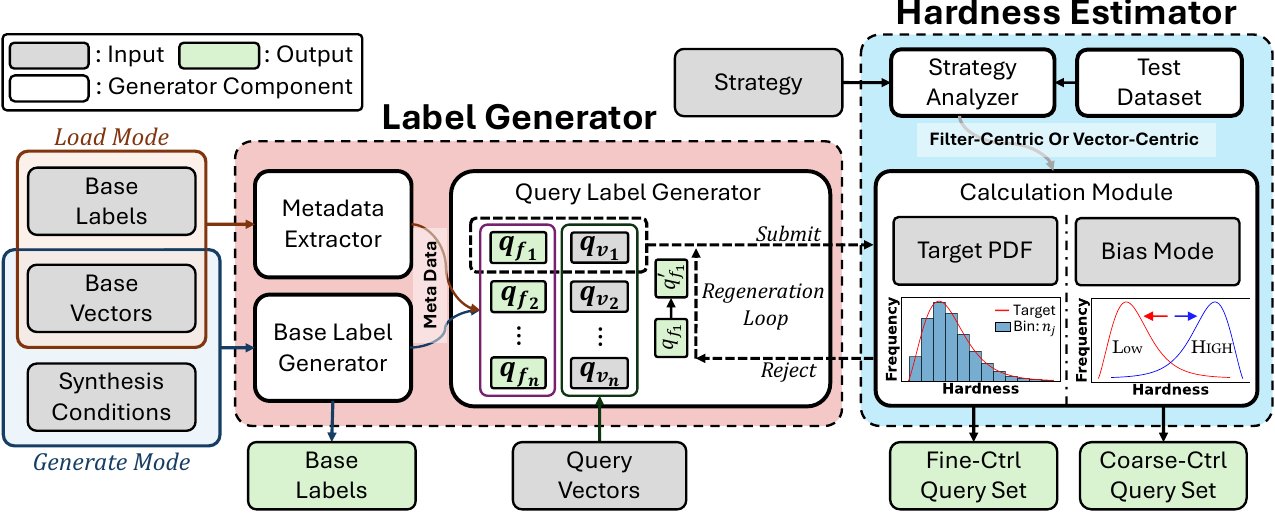}
    \caption{Overview of the hardness-controlled benchmark generator (\HCBGen). The generator supports both {Load} and {Generate} modes for constructing the base data, and controls query-level hardness using either a target hardness profile or coarse-grained bias modes (\textsc{High}, \textsc{Low}, \textsc{Random}).}
    \label{fig:generator}
\end{figure}

\section{Benchmark Generator}

Because real-world hybrid workloads are rarely shareable, FANN evaluations often rely on synthetic benchmarks, whose hidden design choices can bias rankings and obscure robustness. We present \HCBGen, which uses \Hardness~ to synthesize hybrid workloads with controllable, interpretable, and strategy-consistent difficulty.
The generator supports \reviewbox{R1\\R4}\ra{(i) reproducible, hardness-neutral fixed benchmarks for \emph{fair comparison}, }(ii) \emph{stress testing} by emphasizing hard queries, and (iii) \emph{workload approximation} by matching a hardness profile that can be shared without exposing raw query logs.

\subsection{Architecture}
\label{sec:overview}
Figure~\ref{fig:generator} separates the generator into a \emph{Label Generator} and a \emph{Hardness Estimator}. The label generator proposes candidate queries, while the hardness estimator enforces hardness-aware acceptance through a regeneration loop. This separation allows hardness control to remain strategy-consistent across heterogeneous indices.

\begin{algorithm}[t]
\footnotesize
\caption{Coarse-Grained Hardness-Controlled Query Generation (\textsc{High/Low/Random}).}
\label{alg:coarse_hardness}
\begin{algorithmic}[1]
\REQUIRE Base vectors $V$, base labels $L$, mode $\mathcal{M}$, target size $N$, Budget $B$
\ENSURE Query set $\mathcal{Q}$ with $|\mathcal{Q}|=N$. 

\STATE Sample an initial query pool and compute hardness values
\IF{$\mathcal{M}\in\{\textsc{High},\textsc{Low}\}$}
\STATE Set threshold $\tau \leftarrow \mathrm{perc}_{80}(H)$ for \textsc{High}, $\mathrm{perc}_{20}(H)$ for \textsc{Low}
\ENDIF

\FOR{each query vector $q_v$ in the initial pool}
    \STATE Initialize candidate set $\mathcal{C}\leftarrow\emptyset$
    \FOR{$i=1$ to $B$}
        \STATE Sample filter $q_f$ and form $q=(q_v,q_f)$
        \IF{$\mathcal{M}=\textsc{Random}$ \OR $q$ satisfies $\tau$}
            \STATE $\mathcal{Q}\leftarrow\mathcal{Q}\cup\{q\}$; \textbf{break}
        \ENDIF
        \STATE $\mathcal{C}\leftarrow\mathcal{C}\cup\{q\}$
    \ENDFOR
    \IF{no query accepted}
        \STATE $\mathcal{Q}\leftarrow\mathcal{Q}\cup\{\arg\min_{q\in\mathcal{C}} |H(q)-\tau|\}$; 
    \ENDIF
\ENDFOR
\STATE \textbf{return} $\mathcal{Q}$
\end{algorithmic}
\end{algorithm}

\subsubsection{Label Generator}

The label generator supports two base data modes—\emph{Load} and \emph{Generate}—because hybrid benchmarking arises in two fundamentally different settings. In one setting, practitioners have a labeled base data (i.e., base vectors with associated base labels), but the corresponding query logs cannot be shared; the community often has only public vector-only corpora, where structured base labels are unavailable and must be synthesized to support hybrid (vector+filter) benchmarking. 
The two modes trade off \emph{semantic fidelity} and \emph{controlled synthesis}: \textsc{Load} preserves the semantics and structure of a given corpus, whereas \textsc{Generate} creates a synthetic label space with explicit control over workload properties.

In {Load} mode, the generator takes base vectors together with their existing base labels and uses a \emph{Metadata Extractor} to infer the structural metadata needed for query construction—such as attribute domains, cardinalities, value frequencies, and missingness patterns. This metadata is forwarded to the \emph{Query Label Generator}, which samples query labels that are {schema-compatible} with the loaded base labels.
As a result, generated predicates respect the original label vocabulary and distributional structure, enabling realistic query synthesis when the original query workload is unavailable.

In {Generate} mode, the generator takes only base vectors and a set of label synthesis conditions, and uses a \emph{Base Label Generator} to synthesize base labels from scratch. The synthesis conditions define the label space and its statistical properties, and also include an explicit control for vector--label alignment.
Concretely, our default synthesizer partitions the vector space into clusters and assigns attribute values in a DLC-controlled manner so that vectors within the same cluster share values with a tunable strength. This is important because vector--label alignment strongly influences the effectiveness of pruning behaviors in hybrid indices, and thus determines the regime in which a benchmark evaluates robustness versus best-case performance. The synthesized labels are then summarized into the same type of metadata as in {Load}, ensuring that subsequent query generation remains structurally consistent.

Given the extracted or synthesized metadata and the query vectors, the \emph{Query Label Generator} samples candidate predicates and forms hybrid queries. 
\reviewbox{R1.R3 R1.D2}\ra{Since most hybrid-native indices primarily support categorical predicates, it generates predicates composed of one or more \texttt{label = value} conditions. 
Because FANN assumes single-table data with vector embeddings, cross-table operations such as joins are out of scope.}
Candidates are then passed to the hardness estimator, and rejected queries are returned for regeneration, forming a hardness-controlled regeneration loop (Section~\ref{sec:hardness_controlled_generation}). This design decouples label construction from hardness control: the label generator defines \textbf{what} predicates are valid and realistic, while the hardness estimator decides \textbf{which} valid queries are retained to satisfy the target hardness regime.

\begin{algorithm}[t]
\footnotesize
\caption{Fine-Grained Hardness-Controlled Query Generation (\textsc{Match-PDF}) with Budgeted Closest-Fill Fallback.}
\label{alg:fine_hardness_stall_relax}
\begin{algorithmic}[1]
\REQUIRE Base vectors $V$, base labels $L$, target hardness PDF $\mathcal{P}$, target size $N$, bin count $B$.
\ENSURE Query set $\mathcal{Q}$ with $|\mathcal{Q}|=N$.

\STATE Initialize $\mathcal{Q}\leftarrow\emptyset$.
\STATE Discretize hardness into $B$ disjoint intervals $\{I_j\}_{j=1}^{B}$ according to $\mathcal{P}$; assign target counts $n_j$.
\STATE Initialize candidate pools $R_j \leftarrow $ empty list for all bins $j$.

\WHILE{$|\mathcal{Q}|<N$}
    \STATE Sample candidate query $q=(q_v,q_f)$; compute hardness $h\gets H(q)$.
    \STATE Find bin $j$ such that $h\in I_j$.
    \IF{$n_j>0$}
        \STATE $\mathcal{Q}\leftarrow\mathcal{Q}\cup\{q\}$; $n_j\leftarrow n_j-1$.
    \ENDIF
    
    \FOR{each bins $j$ with $n_j>0$}
        \STATE \textbf{if} $h \notin I_j$ \textbf{then} store $(\mathrm{dist}(h, I_j), q_i)$ in $R_j$.
    \ENDFOR
    
    \IF{$q$ is the last candidate query}
        \FOR{each bin $j$ with $n_j>0$}
            \STATE fill remaining slots with $\arg\min$-dist unused candidates from $R_j$.
        \ENDFOR
    \ENDIF
\ENDWHILE

\STATE \textbf{return} $\mathcal{Q}$.
\end{algorithmic}
\end{algorithm}

\subsubsection{Hardness Estimator}
\label{sec:hardness_estimator}

Hardness-based control requires strategy awareness because different hybrid indices enforce vector and filter constraints in different orders (Section~\ref{sec:strategy_dependent}). Accordingly, the hardness estimator consists of a \emph{Strategy Analyzer} and a \emph{Hardness Calculation Module}. The strategy analyzer first classifies the target strategy into a pruning family, which fixes the strategy-consistent interpretation of hardness. The hardness calculation module then computes the strategy-conditioned hardness score of each candidate query (Section~\ref{sec:strategy_dependent}) and applies the user-specified control rule to decide whether to keep the query or reject it.


\rb{To \reviewbox{R2\\R1}classify a strategy with minimal overhead, we maintain two fixed probe datasets that were chosen empirically to exhibit the most opposing characteristics across the two pruning families. 
Specifically, these probes are reduced versions of SIFT1M-1 and SIFT1M-7 from Table~\ref{tab:dataset_configurations}, where both the number of base vectors and the number of queries are downsampled by a factor of ten. Given a target strategy, the analyzer builds the index if needed, runs searches on both probes, and records the resulting recall--QPS trade-off curves. It then summarizes each curve using a single Pareto-optimal score computed by the same procedure described in Section~\ref{sec:dataset_eval_method}, and classifies the strategy by comparing the two scores. }


\ra{This estimation pipeline correctly \reviewbox{R2.W3 R1.R2 R1.W3}classifies not only Post-Filtering and Pre-Filtering, but also all hybrid-native FANN strategies, including ACORN, UNG, NHQ, RWalks, and FDANN, in exact agreement with the theoretical categorization in~\cite{lin}. This is also reflected in Figure~\ref{fig:total_heatmap}, where the ordering induced by \Hardness~ remains consistently negative across all strategies.}

\ra{To handle future implementations \reviewbox{R1.W4 R2.R3}that may not show a clear preference for either probe workload, we classify a strategy as \textsc{Mixed} when the two probe scores are too close to distinguish. \textsc{Mixed} is a robust fallback for ambiguous cases: rather than relying on a single family decision, the generator enforces hardness constraints under both interpretations, reducing vulnerability to borderline misclassification. This keeps probing a one-time, cacheable process while preventing classification uncertainty from affecting hardness control.}
\rb{\reviewbox{R2\\D3}Because the index analyzer uses small probe datasets, this step requires at most one minute.}

\begin{table}[t]
\footnotesize
\centering
\setlength{\tabcolsep}{6pt}
\caption{Label-synthesis configurations used in prior hybrid indexing studies, summarizing the number of attributes, attribute cardinalities, value distributions, missing probabilities, and data–label correlation (DLC).}
\label{tab:dataset_previous}
\begin{adjustbox}{max width=1\columnwidth}
\begin{tabular}{lccccc}
\toprule
Method & \#Attr & Cardi. & Distribution & Missing Prob. & DLC \\
\midrule
NHQ    & 3 & $(3,2,2)$               & Random & 0       & No \\
ACORN  & 1 & $(12)$                  & Random & 0       & No \\
UNG\tablefootnote{UNG synthesizes labels by treating each attribute--value pair as a single label and assigning 12 such labels under a Zipfian distribution. To replicate this, we use 12 attributes of cardinality 1 and implement the Zipfian skew through missing probabilities, yielding a functionally equivalent labeling policy.
}    & 12& $(1,1,\dots 1)$& -- & Zipf& No \\
RWalks & 6 & $(100,20,10,5,3,2)$     & Random & 0       & No \\
\bottomrule
\end{tabular}
\end{adjustbox}
\end{table}

\subsection{Hardness-Controlled Query Generation}
\label{sec:hardness_controlled_generation}
The generator supports two hardness-control families. The first provides coarse-grained modes for stress testing and hardness-neutral fixed benchmarks, while the second enables fine-grained profile matching for workload approximation. Both use the same accept/reject loop and differ only in the acceptance rule.

\subsubsection{Coarse-Grained Bias Modes}
The coarse-grained modes bias the workload toward hard or easy queries (Algorithm~\ref{alg:coarse_hardness}). Rather than using a fixed, absolute hardness threshold, which can vary widely across datasets and label constructions, we derive a cutoff adaptively from a small initial sample of generated queries. For example, \textsc{High} uses a cutoff corresponding to the top portion of the sampled \Hardness, while \textsc{Low} uses a cutoff corresponding to the bottom portion. A candidate query is accepted if its \Hardness~ falls into the desired region (hard for \textsc{High}, easy for \textsc{Low}); \textsc{Random} disables hardness-based selection and accepts randomly generated queries. \reviewbox{R1\\R4}\ra{\textsc{Random} performs no hardness-based filtering, producing a reproducible, strategy-independent fixed benchmark. Because it retains queries spanning a broad range of hardness values, it yields a benchmark with diverse query difficulty that can be used for fair comparison across hybrid index designs.} To ensure termination and stable generation time, we cap the number of regeneration attempts per query. If no candidate meets the desired condition within this retry limit, we accept the candidate whose difficulty score is closest to the cutoff, which guarantees completion while still preserving a strong bias.

\subsubsection{Fine-Grained Target-Profile Matching}
Fine-grained control targets a user-specified hardness profile to approximate an unknown real workload using \Hardness~ as a shareable signal (Algorithm~\ref{alg:fine_hardness_stall_relax}). Concretely, the generator first partitions the difficulty range into a fixed number of intervals and assigns a target number of queries to each interval according to the desired profile.
During generation, each sampled candidate query is scored and placed into the corresponding interval; it is accepted only if that interval still needs more queries, and otherwise it is rejected and regeneration continues. To keep the procedure practical, we enforce a finite sampling budget. If generation stalls because some intervals are extremely unlikely under the current base data, we trigger explicit fallback mechanisms, such as filling the remaining slots with the closest available candidates.  We validate this behavior in Section~\ref{sec:fidelity_experiment} by comparing both the similarity of the resulting difficulty profiles and the downstream search behavior.

\begin{figure}[t]
    \centering
    \includegraphics[width=1\linewidth]{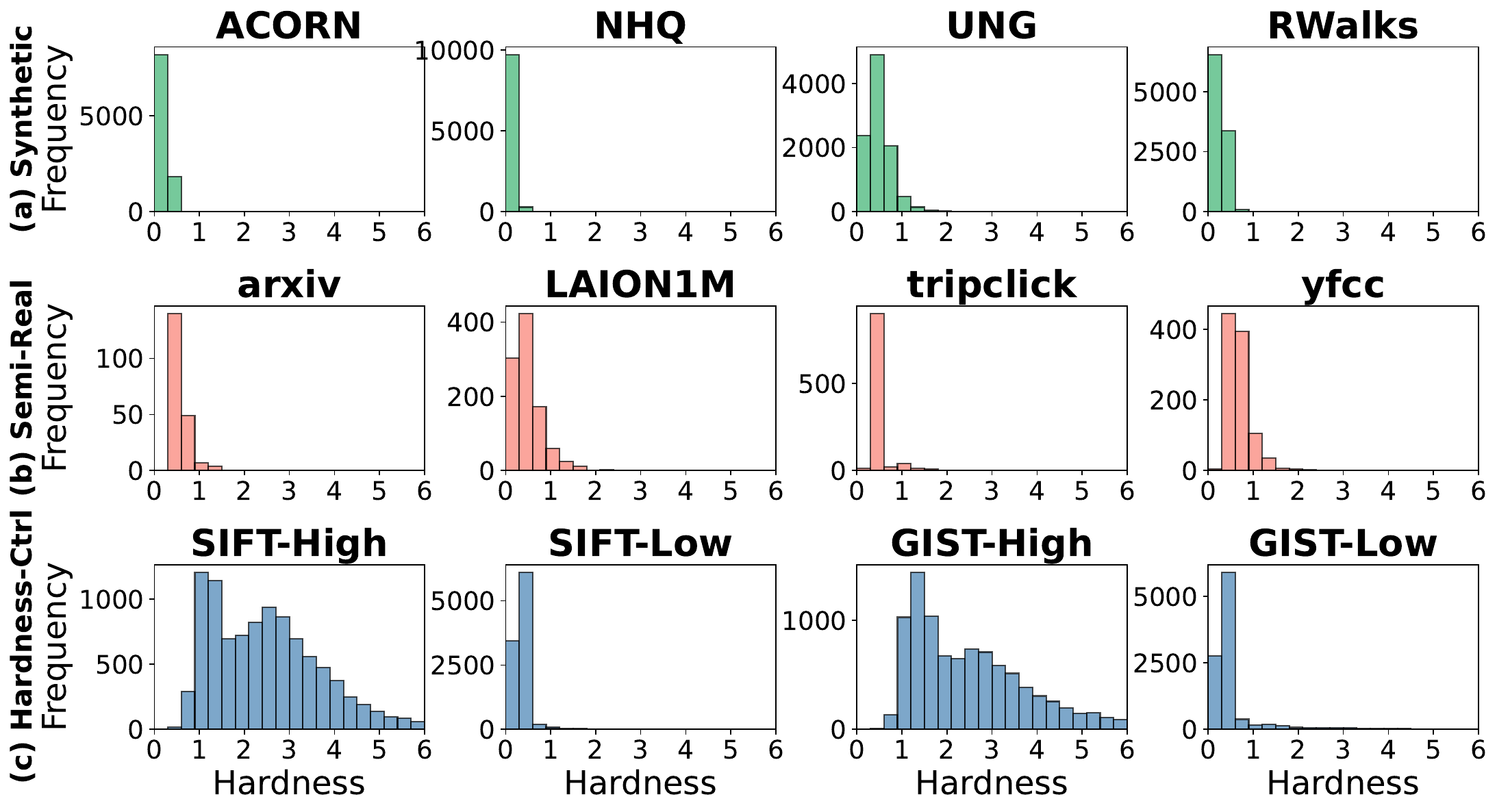}
    \caption{Comparison of \Hardness~ distribution coverage across three workload classes: (a) synthetic workloads adopted from prior studies (SIFT1M-based), (b) semi-real workloads, and (c) hardness-controlled workloads generated by our framework.
    }
    \label{fig:hardness_distribution}
\end{figure}

\begin{figure*}[t]
    \centering
    \includegraphics[width=0.95\linewidth]{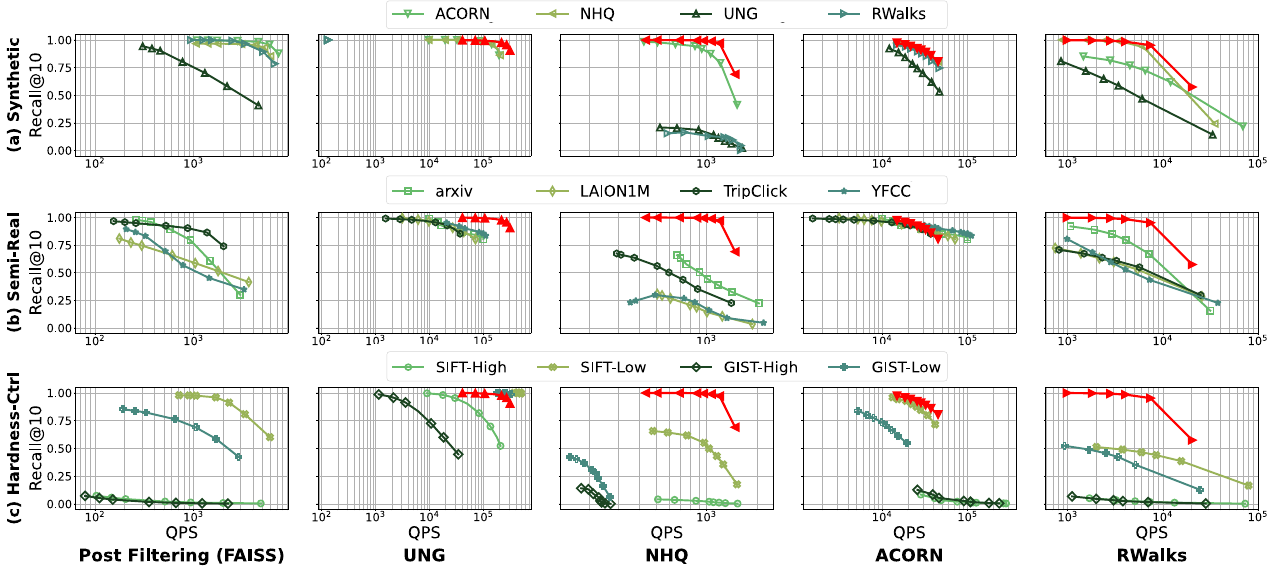}
    \caption{Recall–QPS trade-off curves for five hybrid query processing strategies (Post-Filtering and four hybrid-native indices) across 12 workloads spanning three workload families: (i) synthetic workloads adopted in prior hybrid indexing studies, (ii) semi-real workloads, and (iii) hardness-controlled workloads generated by our method. Red markers overlay the performance reported in each strategy’s original paper (when available), providing a reference anchor for cross-workload comparison.}
    \label{fig:over_estimate}
\end{figure*}

\section{Evaluation}
\label{sec:Experiments}


This section evaluates the proposed \HCBGen from three perspectives: coverage, controllability, and workload approximation. 
We investigate the following research questions:
\begin{itemize}[leftmargin=1.2em]
    \item \textbf{RQ1 (Hardness coverage):} How much of the hardness spectrum is covered by workloads used in prior FANN studies and by semi-real workloads?
    \item \textbf{RQ2 (Hardness controllability):} Can our generator reliably shift the workload toward easy vs.\ hard regimes under identical synthesis settings?
    \item \textbf{RQ3 (Robustness under controlled regimes):} How do representative strategies behave when evaluated across matched, systematically varied hardness regimes?
    \item \textbf{RQ4 (Hardness as a bridge):} If only a hardness profile can be shared, can the generator produce proxy workloads that induce similar performance trends?
\end{itemize}

\subsection{Experimental Setup}
\label{sec:experiments_setting}
We evaluate on three classes of hybrid workloads: (i) synthetic workloads adopted in prior hybrid indexing studies (NHQ, ACORN, UNG, and RWalks~\cite{nhq, acorn, ung, rwalks}), (ii) four semi-real datasets, typically constructed by augmenting a vector-only dataset with synthesized labels, used in prior work (arxiv, LAION1M, tripclick, and yfcc~\cite{arxiv, laion, tripclick, yfcc}), and (iii) four hardness-controlled benchmarks generated by our framework. We follow the original synthesis procedures and parameter recommendations to ensure that our comparisons isolate the effect of \emph{workload design}, rather than retuning or re-implementing indices.
\reviewbox{R1.R3 R1.D2}\ra{In all cases, label synthesis uses predicates composed of categorical conditions of the form \texttt{label=value}, combined with conjunctions.}

For performance benchmarking, we report recall--QPS trade-offs following the parameter sweeps recommended by the corresponding papers to ensure fair and representative evaluation. 
All experiments run on a single Amazon EC2 instance (c5a.8xlarge; 32\,vCPUs\,@\,3.3\,GHz; 64\,GB\,RAM).

For synthetic workloads from prior studies, we fix the base vector dataset to SIFT1M~\cite{sift} and vary only the label/query synthesis policy, so that differences are attributable to workload construction. Label synthesis configurations are summarized in Table~\ref{tab:dataset_previous} and are adopted from the original papers. Query workloads follow the original generation procedures (e.g., single-predicate queries for RWalks and no missing-value constraint on NHQ query filters).
The four semi-real benchmarks are widely used benchmarks in prior FANN studies~\cite{ung, acorn, rwalks}.
They span diverse vector dimensionalities, and attribute configurations~\cite{shi}.

Using the proposed generator, we construct four representative hardness-controlled hybrid benchmarks.
These benchmarks are based on two widely used vector datasets, SIFT1M~\cite{sift} and GIST1M~\cite{sift}, whose vector dimensionalities are 128 and 960, respectively, chosen because SIFT1M has served as a standard baseline across prior hybrid indexing studies, whereas GIST1M offers a markedly higher-dimensional space that stresses vector-centric behavior.
For each base data, we synthesize 3 attributes with a cardinality of 12 per attribute, following a Zipfian distribution with skewness parameter 1.5, a missing probability of 0.5, and 0.0 DLC (data-labal correlation). 
To isolate the effect of hardness control, we generate two workload variants under identical synthesis settings using the coarse-grained \textsc{Low} and \textsc{High} modes (Section~\ref{sec:hardness_controlled_generation}).
This yields \textbf{SIFT-Low}, \textbf{SIFT-High}, \textbf{GIST-Low}, and \textbf{GIST-High}.

\subsection{Hardness Coverage and Control}

We first examine how query hardness is distributed across different workloads and whether our generator enables systematic control.

\subsubsection{Coverage Comparison (RQ1)}

Figure~\ref{fig:hardness_distribution} compares \Hardness \\ distributions. Across these workloads, hardness is typically heavy-tailed: many queries fall into an easy regime with a smaller hard tail. 
Prior synthetic workloads, notably those used by \acorn~and \nhq, are concentrated toward the easy end of the hardness spectrum, and the semi-real workloads similarly occupy a narrow easy range with little mass in harder regimes.
In contrast, our \textbf{SIFT-High} and \textbf{GIST-High} benchmarks span a substantially wider hardness range, including a visible fraction of queries with hardness $>1$ and a hard tail extending beyond 6. 
This broader coverage is important because it enables stress-testing and robustness evaluation in challenging regimes that are weakly represented in commonly used benchmarks, reducing the risk that conclusions hinge on an implicitly ``mild'' workload slice.

\subsubsection{Hardness-Aware Generation (RQ2)}

To verify controllability, we compare \textsc{Low} vs.\ \textsc{High} workloads generated under identical synthesis settings. As shown in Figure~\ref{fig:hardness_distribution}(c), \textsc{Low} concentrates queries in the $[0,1]$ hardness range, whereas \textsc{High} shifts the mass sharply toward harder queries (over 99\% of queries have hardness $>1$).
Crucially, this distributional shift translates into empirical behavior: Figure~\ref{fig:over_estimate}(c) shows that \textsc{Low} workloads yield consistently better recall--QPS trade-offs, while \textsc{High} workloads shift curves toward lower recall under the same search budgets across all evaluated strategies. This validates that hardness is not merely a descriptive statistic: it is an actionable control signal that induces predictable differences in measured performance.

\begin{table}[t]
\centering
\footnotesize
\setlength{\tabcolsep}{5pt}
\renewcommand{\arraystretch}{1.15}
\caption{\rc{Indexing time and memory footprint of four methods on three different synthetic datasets.}}
\begin{adjustbox}{max width=1\columnwidth}
\begin{tabular}{l|ccc|ccc}
\toprule
& \multicolumn{3}{c|}{Indexing Time (seconds)} & \multicolumn{3}{c}{Memory Footprint (MB)} \\
Method 
& \begin{tabular}[c]{@{}c@{}}ACORN Syn.\\[-3pt]{\scriptsize(Simple)}\end{tabular}
& \begin{tabular}[c]{@{}c@{}}UNG Syn.\\[-3pt]{\scriptsize(Medium)}\end{tabular}
& \begin{tabular}[c]{@{}c@{}}RWalks Syn.\\[-3pt]{\scriptsize(Complex)}\end{tabular}
& \begin{tabular}[c]{@{}c@{}}ACORN Syn.\\[-3pt]{\scriptsize(Simple)}\end{tabular}
& \begin{tabular}[c]{@{}c@{}}UNG Syn.\\[-3pt]{\scriptsize(Medium)}\end{tabular}
& \begin{tabular}[c]{@{}c@{}}RWalks Syn.\\[-3pt]{\scriptsize(Complex)}\end{tabular} \\
\midrule
NHQ    & 56  & 40  & 23  & 684  & 695  & 689  \\
ACORN  & 659 & 657 & 685 & 1202 & 1202 & 1202 \\
UNG    & 38  & 40  & 281 & 695  & 693  & 731  \\
RWalks & 342 & 344 & 348 & 1110 & 1203 & 2073 \\
\bottomrule
\end{tabular}
\label{tab:index_build_time_footprint}
\end{adjustbox}
\end{table}

\subsection{Strategy Robustness (RQ3)}
\label{sec:benchmarking_indices}

We next revisit the performance \rc{and indexing / maintenance overhead} of representative hybrid query processing strategies under a common workload suite. 
A key challenge in interpreting the FANN literature is that each method is often evaluated on a workload designed with a method-specific label/query synthesis policy (Table~\ref{tab:dataset_previous}), which complicates cross-paper comparison. Our goal here is not to claim any prior result is ``wrong''; rather, we assess \emph{sensitivity}: whether conclusions drawn under a narrow workload slice remain stable when the hardness regime is varied in a controlled way.

As shown in Figure~\ref{fig:over_estimate}, strategy rankings and trade-off shapes vary substantially across workloads. 
In particular, results reported in the original papers tend to align with easier regions of our workload space, suggesting that performance claims can be workload-sensitive and that robustness under harder regimes is not well captured by commonly used benchmarks. This sensitivity is most evident on \textsc{High} workloads, where recall drops sharply for most strategies, often below 0.2, consistent with the known difficulty of extremely low-selectivity regimes (e.g., around or below $0.1\%$) in which hybrid indices struggle to find enough filter-satisfying candidates~\cite{rwalks}.
\reviewbox{R3.R4 R3.E5}\rc{We also report indexing time and memory footprint in Table~\ref{tab:index_build_time_footprint}. These results are measured on the three synthetic datasets in Table~\ref{tab:dataset_previous}, all built on the same SIFT1M base vectors. Since index construction is independent of the query workload, these costs are driven mainly by the base data label synthesis policy rather than by the \Hardness. Overall, the table shows that indexing overhead is sensitive to label organization: some methods remain stable across datasets, whereas others vary despite sharing the same base vectors.}

\subsubsection{Strategy-Specific Behavior Analysis}
\label{sec:index_specific_analysis}

While all methods degrade under hard workloads, the severity and underlying causes of degradation differ substantially across index designs.

\noindent\textbf{Post Filtering}
 relies exclusively on over-fetched vector candidates to satisfy filter predicates.
Under high-hardness workloads, increasing the over-fetch factor has limited effect, as filter-satisfying candidates remain rare, leading to consistently low recall.
As shown in Figures~\ref{fig:hardness_distribution} and~\ref{fig:over_estimate}, higher \Hardness~ directly translates to poorer search performance across datasets.

\begin{figure}[t]
    \centering
    \includegraphics[width=1\linewidth]{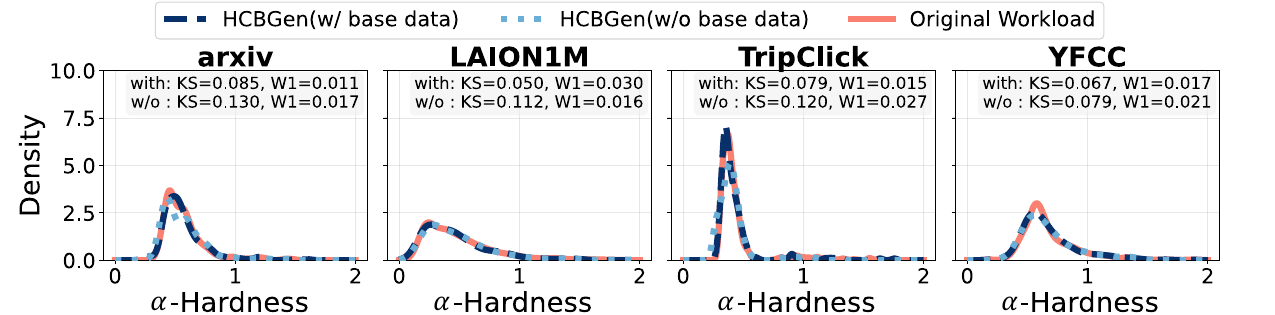}

    \caption{\Hardness~ density on four semi-real datasets, comparing the original workload with two MATCH-PDF workloads generated by HCBGen to match the given distribution, with and without base data. Insets report KS and W1 distances to the original workload (lower is better).}
    \label{fig:fidelity_hardness}
\end{figure}

\begin{figure*}[t]
    \centering
    \includegraphics[width=\linewidth]{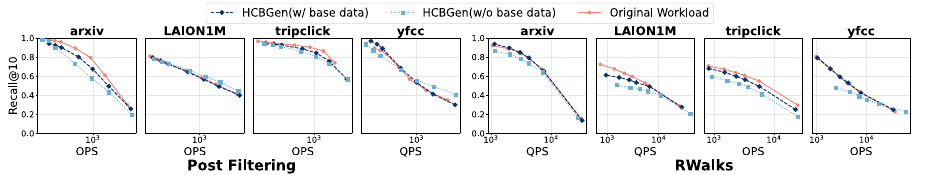}
    
    \caption{Recall–QPS trade-offs on four semi-real datasets for Post-Filtering and RWalks, comparing the original workload with two MATCH-PDF workloads generated by our generator with and without base data. The generated workloads largely preserve the performance trends of the original workload, including the workload generated without base data.}
    
    \label{fig:fidelity_search}
\end{figure*}

\noindent\textbf{UNG}
 is one of the most robust methods to dataset variation among those evaluated. Even extremely low-selectivity queries targeting rare labels, such as \textbf{SIFT-High} in Figure~\ref{fig:over_estimate}(c), can still maintain robust performance because sufficient graph connectivity is preserved, albeit at the cost of substantially higher index construction overhead. 
\reviewbox{R3.R4 R3.E5}\rc{As shown in Table~\ref{tab:index_build_time_footprint}, this design requires UNG to preserve connectivity for rare labels during indexing by partitioning the data into smaller groups, causing both indexing time and memory overhead to increase as the label distribution of the base data becomes more complex.}
These smaller groups can also improve search efficiency. 
However, in workloads such as \rwalks~ with single-label filters, UNG must traverse a large number of groups, which results in lower QPS than in other workloads.


\noindent\textbf{NHQ}
incorporates attribute similarity by averaging binary matches across attributes.
As the number of attributes grows, each match contributes less, reducing recall under complex, high-cardinality labels.
\rc{Interestingly, indexing time decreases as the number of attributes grows, suggesting that index construction may terminate early when too few base vectors share a label to form sufficient edges, resulting in weak graph connectivity and lower recall.}
This pattern is reflected in the recall drops on \ung, \rwalks, and YFCC in Figure~\ref{fig:over_estimate}-NHQ.
Consequently, NHQ is the least robust strategy under hard workloads.

\noindent\textbf{ACORN} evaluates filter satisfaction before traversal and expands only filter-qualified nodes, achieving strong performance under most workloads. 
However, its denser-HNSW strategy breaks down under high-hardness workloads, especially when selectivity falls below approximately $0.1\%$. 
In this regime, dense connectivity alone cannot preserve meaningful connectivity among filter-satisfying nodes, causing the search to terminate prematurely before reaching the true targets and thus inflating QPS despite failed retrieval. 
This behavior is especially evident in \textbf{SIFT-High} and \textbf{GIST-High}. 
\rc{Because this denser-HNSW strategy is predicate-agnostic, its indexing cost is largely insensitive to the base data label distribution, but it consistently requires the highest indexing time and memory footprint among the evaluated methods.}

\noindent\textbf{RWalks} requires base labels to be flattened into binary vectors, making it sensitive to attribute count and missing probability and causing greater performance variation than in other strategies.
RWalks also degrades under high-hardness workloads, consistent with its original finding that recall collapses at extremely low selectivity when index construction fails to establish effective traversal paths.
\rc{Its binary label representation also increases memory overhead as the number of base labels grows, as in \rwalks.}
However, this design is effective when there are few query filters and remains reliable even with large attribute cardinalities.
This behavior aligns with the workload configuration of \rwalks, enabling strong performance with few active labels and high-cardinality attributes.

\subsection{Hardness as a Bridge (RQ4)}
\label{sec:fidelity_experiment}


Another practical implication of \Hardness~ and \HCBGen is that they bridge research benchmarks and real-world hybrid workloads. Because such workloads are rarely released or standardized, we conduct a fidelity experiment by emulating the query workloads of four semi-real datasets. We consider two deployment scenarios. In the first, the base data is available but the original query workload is not, so the generator receives only the target hardness information; this corresponds to the \emph{base labels loading} configuration with \textsc{Match-PDF} mode (\emph{Generated with base data}). In the second, more challenging scenario, neither base data nor query logs are available, and the generator again receives only the target hardness information (\emph{Generated w/o base data}). In this case, base labels are synthesized using a fixed configuration with 2 attributes of cardinality 12. This extreme but practical setting demonstrates the generality of hardness as a standalone workload descriptor.

Figure~\ref{fig:fidelity_hardness} compares the hardness distributions of the original workloads (given) with those of the two proxy workloads generated by our method. 
\textsc{Match-PDF} enables the generator to closely fit the target hardness distribution across all four datasets.
To quantify distributional similarity, we report (i) the \emph{Kolmogorov--Smirnov statistic}\cite{ks} and (ii) the \emph{1-Wasserstein distance}\cite{w1} between the generated and given hardness distributions.
Both metrics consistently show that the generated workloads closely match the given ones.


We further test this argument by asking whether matching hardness distributions also matches performance. Figure~\ref{fig:fidelity_search} reports recall--QPS trade-offs for Post-Filtering and RWalks under three workloads per dataset. The generated workloads closely track the trade-off curves of the given workloads. Notably, even when the generator has no access to the base data, it still reproduces similar performance trends. This supports two conclusions: (i) \Hardness~ is a strong proxy for performance-relevant workload difficulty, and (ii) hardness-controlled generation can faithfully emulate unseen workloads using only hardness profiles.
These results show that our approach can generate privacy-preserving proxy benchmarks that retain the performance-critical properties of real-world hybrid workloads.
This enables realistic benchmarking of hybrid search systems without access to sensitive user logs or proprietary query traces.
More broadly, the results suggest that \Hardness~ captures the dominant factors governing hybrid search performance, making it an effective surrogate for approximating unseen workloads.

\section{Related Works}
\noindent\textbf{Vector Search Hardness Estimation.}
Prior work on vector search has characterized query hardness using intrinsic data properties such as intrinsic dimensionality, local density variation, and hubness, which influence nearest-neighbor structure and distance concentration in high-dimensional spaces~\cite{beyer1999nearest,radovanovic2010hubs,houle2017local}.
More recently, Steiner Hardness provides an execution-oriented notion of query difficulty for graph-based ANN search by estimating the minimum traversal effort required to reach the target neighborhood~\cite{wang2024steiner}.
\reviewbox{R3.R1 R3.W1}\rc{
Our work complements this line by treating \Hardness~ as a broader hybrid extension of Steiner-Hardness.
Specifically, \Hardness~ generalizes the Steiner-Hardness view from vector-only ANN to FANN by retaining graph-traversal difficulty as the fetch component while additionally modeling the over-fetching and filter-induced execution effects unique to hybrid search.
}
To the best of our knowledge, no prior work has explicitly defined FANN query hardness.

\noindent\textbf{FANN Benchmarking.}
Recent work has also introduced FANN benchmarking studies. Shi et al.~\cite{shi} construct semi-real datasets under controlled settings such as label length, selectivity, and dataset size, but still generate queries by sampling labels from the dataset; under our hardness estimator, these queries are biased toward easy cases (Figure~\ref{fig:hardness_distribution}). Lin et al.~\cite{lin} discuss possible query-difficulty factors conceptually, but still rely on query sampling, inheriting label-distribution bias and limiting robustness evaluation. Iff et al.~\cite{iff} introduce a semi-real hybrid dataset with diverse query labels, yet without an explicit notion of query-level difficulty, the representativeness of those queries remains unclear. 
Existing FANN benchmarks therefore lack an explicit notion of query-level hardness, leaving robustness under hard queries unexplored.

\noindent\rb{\textbf{Benchmark Sensitivity in Database Systems.}
\reviewbox{R2\\R2}Benchmark brittleness is a recognized problem across database research. In relational systems, TPC-H and TPC-DS have been shown to miss real-world workloads in schema type, expression complexity, and operator structure~\cite{RDB_TPC}, while cloud-scale production workloads exhibit write-heavy pipelines and long-tailed distributions absent from standard benchmarks~\cite{RDB_cloud}. In ANN search, algorithm rankings shift substantially across datasets~\cite{ANN_Hydra} and also depend on whether queries are easy or hard~\cite{ANN_sota}. Together, these results show that benchmark design choices often determine performance rankings.}

\section{Conclusion}

In this paper, we presented \Hardness, an execution-driven measure of hybrid-query difficulty that aligns consistently with observed FANN performance across datasets and strategies, unlike selectivity- or correlation-based proxies. Using \Hardness~ as a control knob, \HCBGen~ generates hardness-controlled benchmarks and can approximate real workloads by matching hardness profiles, revealing that many existing benchmarks are biased toward easy queries and can overstate robustness.

\begin{acks}
 This work was supported by the [...] Research Fund of [...] (Number [...]). Additional funding was provided by [...] and [...]. We also thank [...] for contributing [...].
\end{acks}


\bibliographystyle{ACM-Reference-Format}
\bibliography{reference}

\end{document}